\documentclass[12pt]{article}
\usepackage{amsfonts,amsmath,amssymb}
\usepackage{color}
\def\C{\mathbb{C}}

\def\Z{\mathbb{Z}}
\def\D{\mathbb{D}}

\def\bq{ \begin{equation} }
\def\eq{ \end{equation} }
\def\ben{ \begin{eqnarray} }
\def\en{ \end{eqnarray} }

\def\frac#1#2{{#1\over #2}}

\def\on#1#2{\mathop{\vbox{\ialign{##\crcr\noalign{\kern2pt}
$\scriptstyle{#2}$\crcr\noalign{\kern2pt\nointerlineskip}
\kern-2pt$\hfil\displaystyle{#1}\hfil$\crcr}}}\limits}

\begin{document}

\baselineskip=15pt
\vspace{1cm} \centerline{\LARGE  \textbf{A simple construction of integrable
}}
\vskip0.4cm
\centerline{\LARGE  \textbf{ Whitham type hierarchies }}

\vskip1cm \hfill
\begin{minipage}{13.5cm}
\baselineskip=10pt
{\large \bf
  A. Odesskii ${}^{}$} \\ [2ex]
{\footnotesize
${}^{}$ Brock University, St. Catharines, Canada }\\

\vskip1cm{\bf Abstract}
\bigskip

A simple construction of Whitham type hierarchies in all genera is suggested. Potentials of these hierarchies are written as integrals of hypergeometric type. Possible generalization 
for universal moduli space is also briefly discussed.

\end{minipage}

\vglue1cm \textbf{Address}:
Brock University,  Niagara Region,  500 Glenridge Ave., St. Catharines, Ont., L2S 3A1 Canada

\textbf{E-mail}:
aodesski@brocku.ca 

\newpage

\tableofcontents

\newpage

\section{Introduction}

The goal of this paper is to give a simple construction for a wide class of integrable quasi-linear systems of the form
\begin{equation}\label{hydr1}
\sum_{l=1}^n\Big(a_{rl}(u_1,...,u_n)\frac{\partial u_l}{\partial t}+b_{rl}(u_1,...,u_n)\frac{\partial u_l}{\partial x}
+c_{rl}(u_1,...,u_n)\frac{\partial u_l}{\partial y}\Big)=0,~r=1,...,m
\end{equation}
where $t,x,y$ are independent variables and $u_1,...,u_n$ are dependent variables. By integrability of a system (\ref{hydr1}) we mean the existence 
of the so-called pseudo-potential representation
\begin{equation}\label{ps}
\frac{\partial\psi}{\partial x}=F(\frac{\partial\psi}{\partial t},u_1,...,u_n),~\frac{\partial\psi}{\partial y}=G(\frac{\partial\psi}{\partial t},u_1,...,u_n). 
\end{equation}
In other words, a system (\ref{hydr1}) is integrable if there exist functions $F,G$ such that the compatibility conditions for (\ref{ps}) are equivalent to (\ref{hydr1}). Writing the system 
(\ref{ps}) in parametric form 
$$\frac{\partial\psi}{\partial t}=P_1(z,{\bf u}),~\frac{\partial\psi}{\partial x}=P_2(z,{\bf u}),~\frac{\partial\psi}{\partial y}=P_3(z,{\bf u}),$$
 allowing an arbitrary number of independent variables $t_1=t,t_2=x,t_3=y,t_4,...,t_N$, and writing compatibility conditions in terms of functions $P_i$, we 
obtain the so-called Whitham type hierarchy. An important class 
of such hierarchies associated with the moduli space of Riemann surfaces of genus $g$ with $n$ punctures (the so-called universal Whitham hierarchy) was constructed 
and studied in \cite{kr1,kr2}. The universal Whitham hierarchy is important in the theory of Frobenius manifolds \cite{dub}, matrix models and other areas of mathematics. 
Note that the set of times in 
the universal Whitham hierarchy coincides with a set of meromorphic differentials on a Riemann surface (holomorphic outside punctures), and that the potentials $P_i(z)$ are 
integrals of these differentials.

In the papers \cite{fer1,fer2,fer3} the general theory of quasi-linear systems of the form (\ref{hydr1}) integrable by hydrodynamic reductions was developed and important 
classification results 
were obtained. In particular, in the paper \cite{fer2} the systems with two equations for two unknowns (i.e. $n=m=2$) were characterized by a complicated system of non-linear PDEs for 
coefficients $a_{rl},~b_{rl},~c_{rl}$. Moreover, it was shown in the same paper that integrability by hydrodynamic reductions (in the case $n=m=2$) is equivalent to existence of a pseudo-potential 
representation. 

In the paper \cite{odes} these systems and their pseudo-potentials were constructed explicitly in terms of arbitrary solutions of a linear system of PDEs of hypergeometric type \cite{gel}
with rational coefficients. 
Moreover, a generalization of this construction to the case of arbitrary $n$ and $m=n$ was done in the same paper. It was clear that the systems constructed in \cite{odes} are associated 
with $\C P^1\setminus\{u_1,...,u_n,0,1,\infty\}$ but constitute a wider class than the universal Whitham hierarchy associated with a rational curve. Further generalization to the case 
of $n+k$ equations with $n$ unknowns (where $0\leq k<n-1$) and to the elliptic case was done in the papers \cite{os1,os2}. Moreover, in was shown in these papers that all 
constructed systems are also integrable by hydrodynamic reductions. It became clear that similar deformations and generalizations of the universal Whitham hierarchy should exist in 
all genera. However, constructions of the papers \cite{odes,os1,os2} were too complicated for direct generalization to Riemann surfaces of genus larger than one. 
Indeed, some expressions for 
derivatives $\frac{\partial P_i}{\partial z}$, $\frac{\partial P_i}{\partial u_j}$ were written down in terms of hypergeometric functions and their derivatives.

Recall that general hypergeometric functions can be constructed 
and studied in two dual ways: as solutions of holonomic linear systems of PDEs and/or as periods of some multiple-valued differential forms.
In this paper by exploring the second method we have solved explicitly the overdetermined systems for $P_i$ found in \cite{odes,os1,os2}, and we write down a simple formula for $P_i$ 
as a single integral 
of hypergeometric type. 
This formula can be easily generalized to all genera. 

Let us describe the contents of the paper. In Section 2 we recall generalities of Whitham type hierarchies. In Section 3 we construct potentials in terms of hypergeometric type 
integrals in the rational case, and in Section 4 we give a similar construction in the elliptic case. In Section 5 we generalize these constructions to higher genus. In Section 6 we 
construct a compatible system of PDEs of hypergeometric type associated with an arbitrary KP tau-function. Some speculations about possible integrable systems 
associated with universal moduli space containing all the Riemann surfaces of finite genus are made and several directions of future research are pointed out.

\section{Whitham type hierarchies}

Given a set of independent variables $t_1,...,t_N$ called times, a set of dependent variables $u_1,...,u_n$ called fields and a set of functions $P_i(z,u_1,...,u_n),~i=1,...,N$ 
called potentials we define a Whitham type hierarchy as compatibility conditions of the following system of PDEs:
\begin{equation}\label{wh}
\frac{\partial\Psi}{\partial t_i}=P_i(z,u_1,...,u_n),~i=1,...,N. 
\end{equation}
Here $\Psi,u_1,...,u_n$ are functions of times $t_1,...,t_N$ and $z$ is a parameter. The system (\ref{wh}) is understood as a parametric way of defining $N-1$ relations between 
partial derivatives $\frac{\partial\Psi}{\partial t_i},~i=1,...,N$ obtained by excluding $z$ from these equations. Assume that the system (\ref{wh}) is compatible. Compatibility 
conditions can be written as 
\begin{equation}\label{comp}
\sum_{l=1}^n\Big(\Big(\frac{\partial P_i}{\partial z}\frac{\partial P_j}{\partial u_l}-
\frac{\partial P_j}{\partial z}\frac{\partial P_i}{\partial u_l}\Big)\frac{\partial u_l}{\partial t_k}+\Big(\frac{\partial P_j}{\partial z}\frac{\partial P_k}{\partial u_l}-
\frac{\partial P_k}{\partial z}\frac{\partial P_j}{\partial u_l}\Big)\frac{\partial u_l}{\partial t_i}+\Big(\frac{\partial P_k}{\partial z}\frac{\partial P_i}{\partial u_l}-
\frac{\partial P_i}{\partial z}\frac{\partial P_k}{\partial u_l}\Big)\frac{\partial u_l}{\partial t_j}\Big)=0 
\end{equation}
where $i,~j,~k=1,...,N$ are pairwise distinct. Let $V_{i,j,k}$ be linear space of functions in $z$ spanned by 
$\frac{\partial P_i}{\partial z}\frac{\partial P_j}{\partial u_l}-
\frac{\partial P_j}{\partial z}\frac{\partial P_i}{\partial u_l},~\frac{\partial P_j}{\partial z}\frac{\partial P_k}{\partial u_l}-
\frac{\partial P_k}{\partial z}\frac{\partial P_j}{\partial u_l},~\frac{\partial P_k}{\partial z}\frac{\partial P_i}{\partial u_l}-
\frac{\partial P_i}{\partial z}\frac{\partial P_k}{\partial u_l},~l=1,...,n$.

{\bf Lemma 1.} Let $V_{i,j,k}$ be finite dimensional and $\dim V_{i,j,k}=m$. Then  (\ref{comp}) is equivalent to a hydrodynamic type system of $m$ linearly independent 
equations of the form
\begin{equation}\label{hydr}
\sum_{l=1}^n\Big(a_{rl}(u_1,...,u_n)\frac{\partial u_l}{\partial t_i}+b_{rl}(u_1,...,u_n)\frac{\partial u_l}{\partial t_j}
+c_{rl}(u_1,...,u_n)\frac{\partial u_l}{\partial t_k}\Big)=0,~r=1,...,m.
\end{equation}

{\bf Proof.} Let $\{S_1(z),...,S_m(z)\}$ be a basis in $V_{i,j,k}$ and 

$\frac{\partial P_i}{\partial z}\frac{\partial P_j}{\partial u_l}-
\frac{\partial P_j}{\partial z}\frac{\partial P_i}{\partial u_l}=\sum_{r=1}^mc_{rl}S_r,~\frac{\partial P_j}{\partial z}\frac{\partial P_k}{\partial u_l}-
\frac{\partial P_k}{\partial z}\frac{\partial P_j}{\partial u_l}=\sum_{r=1}^ma_{rl}S_r,~\frac{\partial P_k}{\partial z}\frac{\partial P_i}{\partial u_l}-
\frac{\partial P_i}{\partial z}\frac{\partial P_k}{\partial u_l}=\sum_{r=1}^mb_{rl}S_r.$ 

Substituting these expressions to (\ref{comp}) and equating to zero coefficients at $S_1,...,S_m$ we obtain (\ref{hydr}).

{\bf Remark 1.} In all known examples of integrable Whitham type hierarchies we have $n\leq m\leq 2n-1$. Therefore, this inequality can be regarded as a criterion of integrability.

{\bf Remark 2.} In the theory of integrable systems of hydrodynamic type the system (\ref{wh}) is often referred to as a pseudo-potential representation of the system (\ref{hydr}).

\section{Genus zero case}

Let $u_1,...,u_n\in\C\setminus\{0,1\}$ be pairwise distinct. Fix real numbers $s_1,...,s_{n+2}$. Define
\begin{equation}\label{g0}
P_{\gamma}(z,u_1,...,u_n)=\int_{\gamma}\frac{1}{z-t}\frac{(z-u_1)^{s_1}...(z-u_n)^{s_n}z^{s_{n+1}}(z-1)^{s_{n+2}}}{(t-u_1)^{s_1}...(t-u_n)^{s_n}t^{s_{n+1}}(t-1)^{s_{n+2}}}dt 
\end{equation}
where $\gamma$ is a cycle in $\C\setminus\{u_1,...,u_n,0,1\}$. Note that $u_1,...,u_n,0,1,\infty$ can be endpoints of $\gamma$ and we assume that the corresponding $s_i$ are small 
enough for convergence of our integral. 

{\bf Proposition 1.} For generic values of $s_1,...,s_{n+2}$ the set of functions $P_{\gamma}(z,u_1,...,u_n)$ defines a Whitham type hierarchy with $n$ fields $u_1,...,u_n$ and $N=n+1$ 
times. Compatibility conditions for this potentials are equivalent to a hydrodynamic type system of the form (\ref{hydr}) with $m=n$ linearly independent equations.

{\bf Proof.} Let $I$ be integrand in (\ref{g0}). Computing  
$\frac{\partial P_{\gamma}}{\partial z}=\int_{\gamma}\frac{\partial I}{\partial z}dt=\int_{\gamma}\Big(\frac{\partial }{\partial z}+\frac{\partial }{\partial t}\Big)Idt$ 
and $\frac{\partial P_{\gamma}}{\partial u_i}=\int_{\gamma}\frac{\partial I}{\partial u_i}dt$ we obtain $\frac{\partial P_{\gamma}}{\partial z}=$
$$
-\int_{\gamma}\Big(\sum_{i=1}^n\frac{s_i}{(z-u_i)(t-u_i)}+\frac{s_{n+1}}{zt}+\frac{s_{n+2}}{(z-1)(t-1)}\Big)
\frac{(z-u_1)^{s_1}...(z-u_n)^{s_n}z^{s_{n+1}}(z-1)^{s_{n+2}}}{(t-u_1)^{s_1}...(t-u_n)^{s_n}t^{s_{n+1}}(t-1)^{s_{n+2}}}dt,$$
$$\frac{\partial P_{\gamma}}{\partial u_i}=
\int_{\gamma}\frac{s_i}{(z-u_i)(t-u_i)}\frac{(z-u_1)^{s_1}...(z-u_n)^{s_n}z^{s_{n+1}}(z-1)^{s_{n+2}}}{(t-u_1)^{s_1}...(t-u_n)^{s_n}t^{s_{n+1}}(t-1)^{s_{n+2}}}dt.$$
These formulas can be written as 
$$
\frac{\partial P_{\gamma}}{\partial z}=\Big(\sum_{i=1}^n\frac{f_{\gamma,i}}{z-u_i}+\frac{f_{\gamma,n+1}}{z}+\frac{f_{\gamma,n+2}}{z-1}\Big)
(z-u_1)^{s_1}...(z-u_n)^{s_n}z^{s_{n+1}}(z-1)^{s_{n+2}},$$
\begin{equation}\label{der0}
\frac{\partial P_{\gamma}}{\partial u_i}=-\frac{f_{\gamma,i}}{z-u_i}(z-u_1)^{s_1}...(z-u_n)^{s_n}z^{s_{n+1}}(z-1)^{s_{n+2}}
\end{equation}
where $f_{\gamma,i}$ are independent of $z$. Note that $f_{\gamma,1}+...+f_{\gamma,n+2}=0$. It is clear from (\ref{der0}) that 
$$\frac{\partial P_{\gamma_1}}{\partial z}\frac{\partial P_{\gamma_2}}{\partial u_l}-
\frac{\partial P_{\gamma_2}}{\partial z}\frac{\partial P_{\gamma_1}}{\partial u_l}=\phi_{\gamma_1,\gamma_2,l}(z)(z-u_1)^{2s_1-1}...(z-u_n)^{2s_n-1}z^{2s_{n+1}-1}(z-1)^{2s_{n+2}-1}$$
where $\phi_{\gamma_1,\gamma_2,l}(z)$ are polynomials in $z$ of degree $n-1$. Therefore, the linear span of these functions is $n$-dimensional and applying Lemma 1 we see that 
compatibility conditions are equivalent to a hydrodynamic type system of the form (\ref{hydr}) with $m=n$ linearly independent equations. It is known that the 
linear space spanned 
by $P_{\gamma}$ is $n+2$-dimensional for generic values of $s_1,...,s_{n+2}$. If $\gamma$ is a small circle around $z$, then $P_{\gamma}$ is a constant. Therefore, 
there are $n+1$ nontrivial times in this hierarchy.

{\bf Remark 3.} Let $\omega=\frac{1}{z-t}\frac{(z-u_1)^{s_1}...(z-u_{n+3})^{s_{n+3}}}{(t-u_1)^{s_1}...(t-u_{n+3})^{s_{n+3}}}dt$. If $s_1+...+s_{n+3}=-1$, then $\omega$ is invariant 
with respect to transformations $t\to\frac{at+b}{ct+d},~z\to\frac{az+b}{cz+d},~u_i\to\frac{au_i+b}{cu_i+d}$. Using these transformations we can send $u_{n+1},u_{n+2},u_{n+3}$ to 
$0,1,\infty$ and obtain integrand of (\ref{g0}). 

{\bf Remark 4.} More general hierarchy can be defined by
\begin{equation}\label{g0def}
P_{\gamma_0,...,\gamma_k}(z,u_1,...,u_n)= 
\end{equation}
$$=\frac{\int\limits_{\gamma_0\times...\times\gamma_k}
\frac{\prod\limits_{0\leq i<j\leq k}(t_i-t_j)\cdot (z-u_1)^{s_1}...(z-u_n)^{s_n}z^{s_{n+1}}(z-1)^{s_{n+2}}}
{\prod\limits_{i=0}^k(z-t_i)(t_i-u_1)^{s_1}...(t_i-u_n)^{s_n}t_i^{s_{n+1}}(t_i-1)^{s_{n+2}}}dt_0\wedge...\wedge t_k}
{\int\limits_{\gamma_1\times...\times\gamma_k}
\frac{\prod\limits_{1\leq i<j\leq k}(t_i-t_j)}
{\prod\limits_{i=1}^k(t_i-u_1)^{s_1}...(t_i-u_n)^{s_n}t_i^{s_{n+1}}(t_i-1)^{s_{n+2}}}dt_1\wedge...\wedge t_k}.$$
Here we fix $\gamma_1,...,\gamma_k$ and vary $\gamma_0$. There are $n$ fields $u_1,...,u_n$ and $n+1-k$ times in this hierarchy. Compatibility conditions are equivalent to a 
system of $n+k$ equations of hydrodynamic type.

{\bf Remark 5.} Yet more general hierarchy can be defined by 
\begin{equation}\label{g0deg}
P_{\gamma_0,...,\gamma_k}(z,{\bf u,v})= 
\end{equation}
$$=\frac{\int\limits_{\gamma_0\times...\times\gamma_k}
\frac{\prod\limits_{0\leq i<j\leq k}(t_i-t_j)\cdot (z-u_1)^{s_1}...(z-u_n)^{s_n}z^{s_{n+1}}(z-1)^{s_{n+2}}e^{\Omega(z)}}
{\prod\limits_{i=0}^k(z-t_i)(t_i-u_1)^{s_1}...(t_i-u_n)^{s_n}t_i^{s_{n+1}}(t_i-1)^{s_{n+2}}e^{\Omega(t_i)}}dt_0\wedge...\wedge t_k}
{\int\limits_{\gamma_1\times...\times\gamma_k}
\frac{\prod\limits_{1\leq i<j\leq k}(t_i-t_j)}
{\prod\limits_{i=1}^k(t_i-u_1)^{s_1}...(t_i-u_n)^{s_n}t_i^{s_{n+1}}(t_i-1)^{s_{n+2}}e^{\Omega(t_i)}}dt_1\wedge...\wedge t_k}$$
where 
$$\Omega(p)=\sum\limits_{i=1}^n\sum\limits_{j=1}^{d_{i}-1}\frac{v_{i,j}}{(p-u_i)^j}+\sum\limits_{j=1}^{d_{n+1}-1}\frac{v_{n+1,j}}{p^j}+
\sum\limits_{j=1}^{d_{n+2}-1}\frac{v_{n+2,j}}{(p-1)^j}+
\sum\limits_{j=1}^{d_{n+3}-1}v_{n+3,j}p^j.$$
Here we fix $\gamma_1,...,\gamma_k$ and vary $\gamma_0$. There are $d_1+...+d_{n+3}$ fields $u_1,...,u_n,v_{i,j}$ and $d_1+...+d_{n+3}+1-k$ times in this hierarchy. 
Compatibility conditions are equivalent to a 
system of $d_1+...+d_{n+3}+k$ equations of hydrodynamic type. In particular, for $k=0$ we have
\begin{equation}\label{g0deg0}
P_{\gamma}(z,u_1,...,u_n)=\int_{\gamma}\frac{1}{z-t}\frac{(z-u_1)^{s_1}...(z-u_n)^{s_n}z^{s_{n+1}}(z-1)^{s_{n+2}}\exp(\Omega(z))}
{(t-u_1)^{s_1}...(t-u_n)^{s_n}t^{s_{n+1}}(t-1)^{s_{n+2}}\exp(\Omega(t))}dt. 
\end{equation}
The numbers $d_1,...,d_{n+3}$ are called multiplicities of $u_1,...,u_n,0,1,\infty$ correspondingly. In particular, if all multiplicities are equal to $1$, then we return to 
potentials given by (\ref{g0def}), (\ref{g0}).

\section{Genus one case}

Let $\Gamma=\{l_1+l_2\tau;~l_1,l_2\in\Z\}\subset\C$ be a lattice in $\C$ spanned by $1$ and $\tau$ where ${\rm Im}~ \tau>0$. Let $\mathcal{E}=\C/\Gamma$ be the corresponding 
elliptic curve. Define theta-function $\theta(z,\tau)$ by
$$\theta(z,\tau)=e^{-\pi iz}\sum_{l\in\Z}(-1)^le^{2\pi i(lz+\frac{l(l-1)}{2}\tau)}.$$
Note that $\theta(z,\tau)$ can be identified with a holomorphic section of a linear bundle on $\mathcal{E}$, the only zero of $\theta(z,\tau)$ modulo $\Gamma$ is at $z=0$ (see \cite{ell} 
for details). 
In the sequel we will omit the second argument of $\theta$ as it always will be equal to $\tau$. The notation $\theta^{\prime}$ is used
for derivative of $\theta$ by the first argument. We will need the following identities:
$$\theta(-z,\tau)=-\theta(z,\tau),~\theta(z+1)=-\theta(z),~\theta(z+\tau)=-e^{-2\pi i(z+\frac{\tau}{2})}\theta(z),
~\frac{\partial\theta}{\partial\tau}=-\frac{i}{4\pi}\theta^{\prime\prime}-\frac{\pi i}{4}\theta,$$
\begin{equation}\label{id}
\end{equation}
$$\frac{\theta^{\prime}(z-t+\eta)}{\theta(z-t+\eta)}-\frac{\theta^{\prime}(\eta)}{\theta(\eta)}+\frac{\theta^{\prime}(t-u)}{\theta(t-u)}-\frac{\theta^{\prime}(z-u)}{\theta(z-u)}=
-\frac{\theta^{\prime}(0)\theta(z-t)\theta(z-u+\eta)\theta(t-u-\eta)}{\theta(\eta)\theta(z-t+\eta)\theta(z-u)\theta(t-u)}.$$

Let $u_1,...,u_n,0\in\C$ be pairwise distinct modulo $\Gamma$. Fix real numbers $s_1,...,s_{n+1}$ such that $s_1+...+s_{n+1}=0$ and complex numbers $a,b$. Let $\eta=s_1u_1+...+s_nu_n+a$. 
Define
\begin{equation}\label{g1}
P_{\gamma}(z,u_1,...,u_n,\tau)=\int_{\gamma}\frac{\theta^{\prime}(0)\theta(z-t+\eta)}{\theta(\eta)\theta(z-t)}\frac{\theta(z-u_1)^{s_1}...\theta(z-u_n)^{s_n}\theta(z)^{s_{n+1}}}
{\theta(t-u_1)^{s_1}...\theta(t-u_n)^{s_n}\theta(t)^{s_{n+1}}}e^{b(z-t)}dt. 
\end{equation}
where $\gamma$ is a cycle in $\C\setminus\{u_1,...,u_n,0\}$. Note that $u_1,...,u_n,0$ can be endpoints of $\gamma$ and we assume that the corresponding $s_i$ are small 
enough for convergence of our integral. 

{\bf Proposition 2.} For generic values of $s_1,...,s_{n+1}$ the set of functions $P_{\gamma}(z,u_1,...,u_n,\tau)$ defines a Whitham type hierarchy with $n+1$ fields $u_1,...,u_n,\tau$ 
and $N=n+1$ times. Compatibility conditions for these potentials are equivalent to a hydrodynamic type system of the form (\ref{hydr}) with $m=n+1$ linearly independent equations.

{\bf Proof.} Let $I$ be integrand in (\ref{g1}). Computing  
$\frac{\partial P_{\gamma}}{\partial z}=\int_{\gamma}\frac{\partial I}{\partial z}dt=\int_{\gamma}\Big(\frac{\partial }{\partial z}+\frac{\partial }{\partial t}\Big)Idt$, 
$\frac{\partial P_{\gamma}}{\partial u_i}=\int_{\gamma}\frac{\partial I}{\partial u_i}dt$,
$\frac{\partial P_{\gamma}}{\partial \tau}=\int_{\gamma}\frac{\partial I}{\partial \tau}dt$ and using (\ref{id}) we obtain 
$\frac{\partial P_{\gamma}}{\partial z}=\frac{\theta^{\prime}(0)^2}{\theta(\eta)^2}\times$
$$
\int_{\gamma}\Big(\sum_{i=1}^n\frac{s_i\theta(z-u_i+\eta)\theta(t-u_i-\eta)}{\theta(z-u_i)\theta(t-u_i)}+
\frac{s_{n+1}\theta(z+\eta)\theta(t-\eta)}{\theta(z)\theta(t)}\Big)
\frac{\theta(z-u_1)^{s_1}...\theta(z-u_n)^{s_n}\theta(z)^{s_{n+1}}e^{bz}}{\theta(t-u_1)^{s_1}...\theta(t-u_n)^{s_n}\theta(t)^{s_{n+1}}e^{bt}}dt,$$
$$\frac{\partial P_{\gamma}}{\partial u_i}=
-\int_{\gamma}\frac{s_i\theta^{\prime}(0)^2\theta(z-u_i+\eta)\theta(t-u_i-\eta)}{\theta(\eta)^2\theta(z-u_i)\theta(t-u_i)}
\frac{\theta(z-u_1)^{s_1}...\theta(z-u_n)^{s_n}\theta(z)^{s_{n+1}}e^{bz}}{\theta(t-u_1)^{s_1}...\theta(t-u_n)^{s_n}\theta(t)^{s_{n+1}}e^{bt}}dt,$$
$\frac{\partial P_{\gamma}}{\partial \tau}=-\frac{\theta^{\prime}(\eta)}{2\pi i\theta(\eta)}\frac{\partial P_{\gamma}}{\partial z}+\frac{\theta^{\prime}(0)^2}{2\pi i\theta(\eta)^2}\times$
$$
\int_{\gamma}\Big(\sum_{i=1}^n\frac{s_i\theta^{\prime}(z-u_i+\eta)\theta(t-u_i-\eta)}{\theta(z-u_i)\theta(t-u_i)}+
\frac{s_{n+1}\theta^{\prime}(z+\eta)\theta(t-\eta)}{\theta(z)\theta(t)}\Big)
\frac{\theta(z-u_1)^{s_1}...\theta(z-u_n)^{s_n}\theta(z)^{s_{n+1}}e^{bz}}{\theta(t-u_1)^{s_1}...\theta(t-u_n)^{s_n}\theta(t)^{s_{n+1}}e^{bt}}dt,$$
These formulas can be written as 
$$
\frac{\partial P_{\gamma}}{\partial z}=\Big(\sum_{i=1}^n\frac{f_{\gamma,i}\theta(z-u_i+\eta)}{\theta(\eta)\theta(z-u_i)}+\frac{f_{\gamma,n+1}\theta(z+\eta)}{\theta(\eta)\theta(z)}\Big)
\theta(z-u_1)^{s_1}...\theta(z-u_n)^{s_n}\theta(z)^{s_{n+1}},$$
\begin{equation}\label{der1}
\frac{\partial P_{\gamma}}{\partial u_i}=-\frac{f_{\gamma,i}\theta(z-u_i+\eta)}{\theta(\eta)\theta(z-u_i)}\theta(z-u_1)^{s_1}...\theta(z-u_n)^{s_n}\theta(z)^{s_{n+1}},
\end{equation}
$$
\frac{\partial P_{\gamma}}{\partial \tau}=-\frac{\theta^{\prime}(\eta)}{2\pi i\theta(\eta)}\frac{\partial P_{\gamma}}{\partial z}+
\Big(\sum_{i=1}^n\frac{f_{\gamma,i}\theta^{\prime}(z-u_i+\eta)}{2\pi i\theta(\eta)\theta(z-u_i)}+\frac{f_{\gamma,n+1}\theta^{\prime}(z+\eta)}{2\pi i\theta(\eta)\theta(z)}\Big)
\theta(z-u_1)^{s_1}...\theta(z-u_n)^{s_n}\theta(z)^{s_{n+1}}$$
where $f_{\gamma,i}$ are independent of $z$. It is clear from (\ref{der1}) that 
$$\frac{\partial P_{\gamma_1}}{\partial z}\frac{\partial P_{\gamma_2}}{\partial u_l}-
\frac{\partial P_{\gamma_2}}{\partial z}\frac{\partial P_{\gamma_1}}{\partial u_l}=\phi_{\gamma_1,\gamma_2,l}(z)\theta(z-u_1)^{2s_1}...\theta(z-u_n)^{2s_n}\theta(z)^{2s_{n+1}},~l=1,...,n,$$
$$\frac{\partial P_{\gamma_1}}{\partial z}\frac{\partial P_{\gamma_2}}{\partial \tau}-
\frac{\partial P_{\gamma_2}}{\partial z}\frac{\partial P_{\gamma_1}}{\partial \tau}=\phi_{\gamma_1,\gamma_2,n+1}(z)\theta(z-u_1)^{2s_1}...\theta(z-u_n)^{2s_n}\theta(z)^{2s_{n+1}}$$
where $\phi_{\gamma_1,\gamma_2,l}(z)$ are meromorphic functions in $z$ with simple poles at $u_1,...,u_n,0$ only. Moreover, these functions satisfy quasi-periodicity properties:
$$\phi_{\gamma_1,\gamma_2,l}(z+1)=\phi_{\gamma_1,\gamma_2,l}(z),~\phi_{\gamma_1,\gamma_2,l}(z+\tau)=e^{-2\pi i\eta}\phi_{\gamma_1,\gamma_2,l}(z),~l=1,...,n+1.$$
 Therefore, the linear span of these functions is $n+1$-dimensional and applying Lemma 1 we see that 
compatibility conditions  are equivalent to a hydrodynamic type system of the form (\ref{hydr}) with $m=n+1$ linearly independent equations. 
The linear space spanned 
by $P_{\gamma}$ is $n+2$-dimensional for generic values of $s_1,...,s_{n+1}$. If $\gamma$ is a small circle around $z$, then $P_{\gamma}$ is a constant. Therefore, 
there are $n+1$ nontrivial times in this hierarchy.

{\bf Remark 6.} Let $\omega=\frac{\theta^{\prime}(0)\theta(z-t+\eta)}{\theta(z-t)}\frac{\theta(z-u_1)^{s_1}...\theta(z-u_n)^{s_n}\theta(z-u_{n+1})^{s_{n+1}}}
{\theta(t-u_1)^{s_1}...\theta(t-u_n)^{s_n}\theta(t-u_{n+1})^{s_{n+1}}}e^{b(z-t)}dt$. If $s_1+...+s_{n+1}=0$, then $\omega$ is invariant 
with respect to simultaneous translations of $z,t,u_1,...,u_{n+1}$. Using these translations we can send $u_{n+1}$ to 
$0$ and obtain integrand of (\ref{g1}). 

{\bf Remark 7.} More general hierarchy can be defined by
\begin{equation}\label{g1def}
P_{\gamma_0,...,\gamma_k}(z,u_1,...,u_n,\tau)=\frac{\theta^{\prime}(0)}{\Delta} \times 
\end{equation}
$$\int\limits_{\gamma_0\times...\times\gamma_k}
\frac{\theta(z-\sum\limits_{i=0}^kt_i+\eta)\prod\limits_{0\leq i<j\leq k}\theta(t_i-t_j)\cdot \theta(z-u_1)^{s_1}...\theta(z-u_n)^{s_n}\theta(z)^{s_{n+1}}e^{bz}}
{\prod\limits_{i=0}^k\theta(z-t_i)\theta(t_i-u_1)^{s_1}...\theta(t_i-u_n)^{s_n}\theta(t_i)^{s_{n+1}}e^{bt_i}}dt_0\wedge...\wedge t_k$$
where
$$\Delta=\int\limits_{\gamma_1\times...\times\gamma_k}
\frac{\theta(\eta-\sum\limits_{i=1}^kt_i)\prod\limits_{1\leq i<j\leq k}\theta(t_i-t_j)}
{\prod\limits_{i=1}^k\theta(t_i-u_1)^{s_1}...\theta(t_i-u_n)^{s_n}\theta(t_i)^{s_{n+1}}e^{bt_i}}dt_1\wedge...\wedge t_k.$$
Here we fix $\gamma_1,...,\gamma_k$ and vary $\gamma_0$. There are $n+1$ fields $u_1,...,u_n,\tau$ and $n+1-k$ times in this hierarchy. Compatibility conditions are equivalent to a 
system of $n+1+k$ equations of hydrodynamic type.

{\bf Remark 8.} Yet more general hierarchy can be defined by 
\begin{equation}\label{g1deg}
P_{\gamma_0,...,\gamma_k}(z,{\bf u,v},\tau)=\frac{\theta^{\prime}(0)}{\Delta} \times
\end{equation}
$$\int\limits_{\gamma_0\times...\times\gamma_k}
\frac{\theta(z-\sum\limits_{i=0}^kt_i+\eta)\prod\limits_{0\leq i<j\leq k}\theta(t_i-t_j)\cdot \theta(z-u_1)^{s_1}...\theta(z-u_n)^{s_n}\theta(z)^{s_{n+1}}e^{bz+\Omega(z)}}
{\prod_{i=0}^k\theta(z-t_i)\theta(t_i-u_1)^{s_1}...\theta(t_i-u_n)^{s_n}\theta(t_i)^{s_{n+1}}e^{bt_i+\Omega(t_i)}}dt_0\wedge...\wedge t_k$$
where 
$$\Delta=\int\limits_{\gamma_1\times...\times\gamma_k}
\frac{\theta(\eta-\sum\limits_{i=1}^kt_i)\prod\limits_{1\leq i<j\leq k}\theta(t_i-t_j)}
{\prod\limits_{i=1}^k\theta(t_i-u_1)^{s_1}...\theta(t_i-u_n)^{s_n}\theta(t_i)^{s_{n+1}}e^{bt_i+\Omega(t_i)}}dt_1\wedge...\wedge t_k,$$
$$\Omega(p)=\sum_{i=1}^n\sum_{j=1}^{d_i-1}v_{i,j}\Omega_j(p-u_i)+\sum_{j=1}^{d_{n+1}-1}v_{n+1,j}\Omega_j(p),~\Omega_j(p)=\frac{\partial^j}{\partial p^j}\log(\theta(p)).$$
Here we fix $\gamma_1,...,\gamma_k$ and vary $\gamma_0$. There are $d_1+...+d_{n+1}+1$ fields $u_1,...,u_n,v_{i,j},\tau$ and $d_1+...+d_{n+1}+1-k$ times in this hierarchy. 
Compatibility conditions are equivalent to a 
system of $d_1+...+d_{n+1}+1+k$ equations of hydrodynamic type. In particular, for $k=0$ we have
\begin{equation}\label{g1deg0}
P_{\gamma}(z,u_1,...,u_n,\tau)=\int_{\gamma}\frac{\theta^{\prime}(0)\theta(z-t+\eta)}{\theta(\eta)\theta(z-t)}\frac{\theta(z-u_1)^{s_1}...\theta(z-u_n)^{s_n}\theta(z)^{s_{n+1}}
e^{bz+\Omega(z)}}
{\theta(t-u_1)^{s_1}...\theta(t-u_n)^{s_n}\theta(t)^{s_{n+1}}e^{bt+\Omega(t)}}dt.
\end{equation}
The numbers $d_1,...,d_{n+1}$ are called multiplicities of $u_1,...,u_n,0$ correspondingly. In particular, if all multiplicities are equal to $1$, then we return to 
potentials given by (\ref{g1def}), (\ref{g1}).

\section{Higher genus case}

Let $\mathcal{E}=\D/\Gamma$ be a compact Riemann surface of genus $g>1$, $\D\subset \C$ its universal covering and $\Gamma=\pi_1(\mathcal{E})$. 
Denote $a_{\alpha},b_{\alpha},~\alpha=1,...,g$ a canonical basis in the homology group $H_1(\mathcal{E},\Z)$. Let us choose a coordinate in $\D$ and use the same symbols for 
holomorphic objects on $\mathcal{E}$ and their lifting on $\D$. Let $\omega_{\alpha}(z)dz$ be the basis of holomorphic 1-forms on 
$\mathcal{E}$ normalized by $\int_{a_{\alpha}}\omega_{\beta}dz=\delta_{\alpha\beta}$. Choose a basepoint $z_0$ and define the Abel map $q_{\alpha}(z)=\int_{z_0}^z\omega(z)dz$. 
Note that $\omega_{\alpha}=q^{\prime}_{\alpha}$. Denote by $E(x,y)(dx)^{-1/2}(dy)^{-1/2}$ the prime form and by 
$$\theta(z_1,...,z_g)=\sum_{{\bf m}\in\Z^g}\exp(2\pi i{\bf m}\cdot{\bf z}+\pi i{\bf m}{\bf B}{\bf m}^t)$$ 
the Riemann theta-function where ${\bf B}=(B_{\alpha\beta})$, $B_{\alpha\beta}= \int_{b_{\alpha}}\omega_{\beta}dz$ is the matrix of $b$-periods. 
See \cite{ell,kok} for details on holomorphic 
objects on Riemann surfaces. Here and in the sequel we use bold 
symbols for the corresponding vectors: ${\bf q}=(q_1,...,q_g),~{\bf z}=(z_1,...,z_g),...$ and ${\bf m}\cdot{\bf z}=m_1z_1+...+m_gz_g$. Recall that 
\begin{equation}\label{id1}
 E(v,u)=-E(u,v),~E(u,v)=u-v+o((u-v)^2),
\end{equation}
$$E(u,v)E(w,t)\theta({\bf z}+{\bf q}(u)+{\bf q}(v))\theta({\bf z}+{\bf q}(w)+{\bf q}(t))+$$$$+E(v,w)E(u,t)\theta({\bf z}+{\bf q}(v)+{\bf q}(w))\theta({\bf z}+{\bf q}(u)+{\bf q}(t))+$$
$$+E(w,u)E(v,t)\theta({\bf z}+{\bf q}(w)+{\bf q}(u))\theta({\bf z}+{\bf q}(v)+{\bf q}(t))=0.$$
The last relation is called Fay identity \cite{fey}. 

Let $u_1,...,u_n\in\D$ be pairwise distinct modulo $\Gamma$. Fix real numbers $s_1,...,s_n$ such that $s_1+...+s_n=1$ and complex vectors ${\bf a,b}\in\C^g$. 
Let $\boldsymbol{\eta}=s_1{\bf q}(u_1)+...+s_n{\bf q}(u_n)+{\bf a}$. 
Define
\begin{equation}\label{g2}
P_{\gamma}(z,u_1,...,u_n)=\int_{\gamma}\frac{\theta({\bf q}(z)-{\bf q}(t)+\boldsymbol{\eta})}{\theta(\boldsymbol{\eta})E(z,t)}\frac{E(z,u_1)^{s_1}...E(z,u_n)^{s_n}}
{E(t,u_1)^{s_1}...E(t,u_n)^{s_n}}e^{{\bf b}\cdot({\bf q}(z)-{\bf q}(t))}dt
\end{equation}
where $\gamma$ is a cycle in $\D\setminus\{u_1,...,u_n\}$. Note that $u_1,...,u_n$ can be endpoints of $\gamma$ and we assume that the corresponding $s_i$ are small 
enough for convergence of our integral. 

{\bf Remark 9.} The function $P_{\gamma}$ does not depend on the choice of coordinate in $\D$. Note that $P_{\gamma}$ is a function of $n+1$ points of $\D$ (with coordinates 
$z,u_1,...,u_n$) and $3g-3$ moduli of a Riemann surface $\mathcal{E}$.

{\bf Proposition 3.} For generic values of $s_1,...,s_n$ the set of functions $P_{\gamma}(z,u_1,...,u_n)$ defines a Whitham type hierarchy with $n+3g-3$ fields ($u_1,...,u_n$ and
 $3g-3$ moduli of $\mathcal{E}$) 
and $N=n+2g-2$ times. Compatibility conditions for these potentials are equivalent to a hydrodynamic type system of the form (\ref{hydr}) with $m=n+3g-3$ linearly independent equations.

Let $I$ be integrand in (\ref{g2}). Computing  
$\frac{\partial P_{\gamma}}{\partial u_i}=\int_{\gamma}\frac{\partial I}{\partial u_i}dt$ and using the Fay identity we obtain 
\begin{equation}\label{der3}
\frac{\partial P_{\gamma}}{\partial u_i}=
\int_{\gamma}\frac{s_i\theta({\bf q}(z)-{\bf q}(u_i)+\boldsymbol\eta)\theta({\bf q}(t)-{\bf q}(u_i)-\boldsymbol{\eta})}{\theta(\boldsymbol{\eta})^2E(z,u_i)E(t,u_i)}
\frac{E(z,u_1)^{s_1}...E(z,u_n)^{s_n}e^{{\bf b}\cdot{\bf q}(z)}}{E(t,u_1)^{s_1}...E(t,u_n)^{s_n}e^{{\bf b}\cdot{\bf q}(t)}}dt.
\end{equation}
Let
\begin{equation}\label{der4}
\frac{\partial P_{\gamma}}{\partial z}=f_{\gamma}(z)
E(z,u_1)^{s_1-1}...E(z,u_n)^{s_n-1}e^{{\bf b}\cdot{\bf q}(z)}.
\end{equation}
One can check that $f_{\gamma}(z)$ is a holomorphic section of a linear bundle of degree $n+3g-3$ on $\mathcal{E}$. Moreover, 
$$f_{\gamma}(u_i)=-\int_{\gamma}\frac{s_i\theta({\bf q}(t)-{\bf q}(u_i)-\boldsymbol{\eta})}{\theta(\boldsymbol{\eta})E(t,u_i)}
\frac{E(u_i,u_1)...\hat{i}...E(u_i,u_n)}{E(t,u_1)^{s_1}...E(t,u_n)^{s_n}e^{{\bf b}\cdot{\bf q}(t)}}dt$$
and, therefore, we have
\begin{equation}\label{der2}
\frac{\partial P_{\gamma}}{\partial u_i}=-\frac{f_{\gamma}(u_i)\theta({\bf q}(z)-{\bf q}(u_i)+\boldsymbol{\eta})}{\theta(\boldsymbol{\eta})E(z,u_i)}
\frac{E(z,u_1)^{s_1}...E(z,u_n)^{s_n}e^{{\bf b}\cdot({\bf q}(z))}}{E(u_i,u_1)...\hat{i}...E(u_i,u_n)}.
\end{equation}
 It is clear from (\ref{der4}), (\ref{der2}) that 
$$\frac{\partial P_{\gamma_1}}{\partial z}\frac{\partial P_{\gamma_2}}{\partial u_l}-
\frac{\partial P_{\gamma_2}}{\partial z}\frac{\partial P_{\gamma_1}}{\partial u_l}=\phi_{\gamma_1,\gamma_2,l}(z)E(z-u_1)^{2s_1-1}...E(z-u_n)^{2s_n-1},~l=1,...,n,$$
where $\phi_{\gamma_1,\gamma_2,l}(z)$ are holomorphic sections of a linear bundle of degree $n+4g-4$ on $\mathcal{E}$.
 Therefore, the linear span of these functions is $n+3g-3$-dimensional and applying Lemma 1 we see that 
compatibility conditions are equivalent to a hydrodynamic type system of the form (\ref{hydr}) with $m=n+3g-3$ linearly independent equations. 
The linear space spanned 
by $P_{\gamma}$ is $n+2g-1$-dimensional for generic values of $s_1,...,s_n$. If $\gamma$ is a small circle around $z$, then $P_{\gamma}$ is a constant. Therefore, 
there are $n+2g-2$ nontrivial times in this hierarchy.

{\bf Remark 10.} More general hierarchy can be defined by
\begin{equation}\label{g2def}
P_{\gamma_0,...,\gamma_k}(z,u_1,...,u_n)= 
\end{equation}
$$=\frac{\int\limits_{\gamma_0\times...\times\gamma_k}
\frac{\theta({\bf q}(z)-\sum\limits_{i=0}^k{\bf q}(t_i)+\boldsymbol{\eta})\prod\limits_{0\leq i<j\leq k}E(t_i,t_j)\cdot E(z,u_1)^{s_1}...E(z,u_n)^{s_n}e^{{\bf b}\cdot{\bf q}(z)}}
{\prod\limits_{i=0}^kE(z,t_i)E(t_i,u_1)^{s_1}...E(t_i,u_n)^{s_n}e^{{\bf b}\cdot{\bf q}(t_i)}}dt_0\wedge...\wedge t_k}
{\int\limits_{\gamma_1\times...\times\gamma_k}
\frac{\theta(\boldsymbol{\eta}-\sum\limits_{i=1}^k{\bf q}(t_i))\prod\limits_{1\leq i<j\leq k}E(t_i,t_j)}
{\prod\limits_{i=1}^kE(t_i,u_1)^{s_1}...E(t_i,u_n)^{s_n}e^{{\bf b}\cdot{\bf q}(t_i)}}dt_1\wedge...\wedge t_k}$$
where $s_1+...+s_n=k+1$.
Here we fix $\gamma_1,...,\gamma_k$ and vary $\gamma_0$. There are $n+3g-3$ fields and $n+2g-2-k$ times in this hierarchy. Compatibility conditions are equivalent to a 
system of $n+3g-3+k$ equations of hydrodynamic type.

{\bf Remark 11.} Yet more general hierarchy can be defined by 
\begin{equation}\label{g2deg}
P_{\gamma_0,...,\gamma_k}(z,{\bf u,v})=
\end{equation}
$$=\frac{\int\limits_{\gamma_0\times...\times\gamma_k}
\frac{\theta({\bf q}(z)-\sum\limits_{i=0}^k{\bf q}(t_i)+\boldsymbol{\eta})\prod\limits_{0\leq i<j\leq k}E(t_i,t_j)\cdot E(z,u_1)^{s_1}...E(z,u_n)^{s_n}e^{{\bf b}\cdot{\bf q}(z)+\Omega(z)}}
{\prod\limits_{i=0}^kE(z,t_i)E(t_i,u_1)^{s_1}...E(t_i,u_n)^{s_n}e^{{\bf b}\cdot{\bf q}(t_i)+\Omega(t_i)}}dt_0\wedge...\wedge t_k}
{\int\limits_{\gamma_1\times...\times\gamma_k}
\frac{\theta(\boldsymbol{\eta}-\sum\limits_{i=1}^k{\bf q}(t_i))\prod\limits_{1\leq i<j\leq k}E(t_i,t_j)}
{\prod\limits_{i=1}^kE(t_i,u_1)^{s_1}...E(t_i,u_n)^{s_n}e^{{\bf b}\cdot{\bf q}(t_i)+\Omega(t_i)}}dt_1\wedge...\wedge t_k}$$
where $s_1+...+s_n=k+1$,
 $$\Omega(p)=\int_{z_0}^p\sum_{i=1}^n\sum_{j=1}^{d_i-1}v_{i,j}\zeta_j(t,u_i)dt,~ \zeta_j(t,u)=\frac{1}{(t-u)^j}+O(1),~\int_{a_{\alpha}}\zeta_j(t,u)dt=0,~\alpha=1,...,g,$$
and $\zeta_j(t,u)$ is holomorphic for $t\ne u$.
Here we fix $\gamma_1,...,\gamma_k$ and vary $\gamma_0$. There are $d_1+...+d_n+3g-3$ fields and $d_1+...+d_n+2g-2-k$ times in this hierarchy. 
Compatibility conditions are equivalent to a 
system of $d_1+...+d_n+3g-3+k$ equations of hydrodynamic type. In particular, for $k=0$ we have
\begin{equation}\label{g2deg0}
P_{\gamma}(z,u_1,...,u_n)=\int_{\gamma}\frac{\theta({\bf q}(z)-{\bf q}(t)+\boldsymbol{\eta})}{\theta(\boldsymbol{\eta})E(z,t)}\frac{E(z,u_1)^{s_1}...E(z,u_n)^{s_n}
e^{{\bf b}\cdot{\bf q}(z)+\Omega(z)}}
{E(t,u_1)^{s_1}...E(t,u_n)^{s_n}e^{{\bf b}\cdot{\bf q}(t)+\Omega(t)}}dt.
\end{equation}
The numbers $d_1,...,d_n$ are called multiplicities of $u_1,...,u_n$ correspondingly. In particular, if all multiplicities are equal to $1$, then we return to 
potentials given by (\ref{g2def}), (\ref{g2}).

\section{Hypergeometric type systems associated with an arbitrary tau-function}

Compatibility conditions for (\ref{der3}) and (\ref{der4}) imply that the functions $f_{\gamma}(z)$ satisfy the following system of PDEs:
\begin{equation}\label{comp1}
\frac{\partial f(z)}{\partial u_i}=-\frac{(s_i-1)\frac{\partial E(z,u_i)}{\partial u_i}}{E(z,u_i)}f(z)- 
\end{equation}
$$-\frac{\theta({\bf q}(z)-{\bf q}(u_i)+\boldsymbol{\eta})E(z,u_1),,,\hat{i}...E(z,u_n)}{\theta(\boldsymbol{\eta})E(u_i,u_1),,,\hat{i}...E(u_i,u_n)}f(u_i)
\Big({\bf b}\cdot {\bf q}^{\prime}(z)+\sum_{j=1}^ns_j\frac{\frac{\partial E(z,u_j)}{\partial z}}{E(z,u_j)}-
$$
$$-\frac{\frac{\partial E(z,u_i)}{\partial z}}{E(z,u_i)}+
\frac{{\bf q}^{\prime}(z)\cdot\boldsymbol{\theta}^{\prime}({\bf q}(z)-{\bf q}(u_i)+\boldsymbol{\eta})}{\theta({\bf q}(z)-{\bf q}(u_i)+\boldsymbol{\eta})}\Big),~i=1,...,n$$
where ${\bf q}^{\prime}(z)\cdot\boldsymbol{\theta}^{\prime}(\boldsymbol{\eta})=\sum_{j=1}^gq^{\prime}_j(z)\frac{\partial\theta(\boldsymbol{\eta})}{\partial\eta_j}$. In particular, 
setting $z=u_j,~j\ne i$ in (\ref{comp1}) and denoting $f_j=\frac{f(u_j)}{E(u_j,u_1)...\hat{j}...E(u_j,u_n)}$ we obtain the following system:
\begin{equation}\label{comp2}
\frac{\partial f_j}{\partial u_i}=-\frac{s_i\frac{\partial E(u_i,u_j)}{\partial u_i}}{E(u_i,u_j)}f_j+
\frac{s_j\theta({\bf q}(u_j)-{\bf q}(u_i)+\boldsymbol{\eta})}{\theta(\boldsymbol{\eta})E(u_i,u_j)}f_i,~i\ne j=1,...,n. 
\end{equation}

{\bf Proposition 4.} Each of the systems (\ref{comp1}), (\ref{comp2}) is compatible by virtue of (\ref{id1}). In other words, let $q_1(z),...,q_g(z),E(x,y),\theta(t_1,...,t_g)$ be 
arbitrary holomorphic functions satisfying (\ref{id1}). Then the system (\ref{comp1}) for a single function $f(z,u_1,...,u_n)$ and the system (\ref{comp2}) for $n$ 
functions $f_i(u_1,...,u_n),~i=1,...,n$ are both compatible. Recall that $\boldsymbol{\eta}=s_1{\bf q}(u_1)+...+s_n{\bf q}(u_n)+{\bf a}$.

Proof is a straightforward computation using (\ref{id1}). 

Let us set $g=\infty,~E(x,y)=x-y,~q_i(z)=\frac{z^i}{i},~i=1,2,...$ and $\theta=\tau$ where $\tau$ is an arbitrary KP tau-function \cite{tau}. Recall that $\tau$ satisfies 
the following Fay type identity:
$$(a-b)(c-d)\tau({\bf t}+[a]+[b])\tau({\bf t}+[c]+[d])+(b-c)(a-d)\tau({\bf t}+[b]+[c])\tau({\bf t}+[a]+[d])+$$
$$+(c-a)(b-d)\tau({\bf t}+[c]+[a])\tau({\bf t}+[b]+[d])=0$$
where ${\bf t}=(t_1,t_2,...)$ and $[a]=(a,\frac{a^2}{2},...)$. The system (\ref{comp2}) takes the form
\begin{equation}\label{comp3}
\frac{\partial f_j}{\partial u_i}=\frac{s_i}{u_j-u_i}f_j+
\frac{s_j\tau([u_j]-[u_i]+\boldsymbol{\eta})}{(u_i-u_j)\tau(\boldsymbol{\eta})}f_i,~i\ne j=1,...,n 
\end{equation}
where $\boldsymbol\eta=s_1[u_1]+...+s_n[u_n]+{\bf a}$. This system is compatible for arbitrary constants $s_1,...,s_n$, $a_1,a_2,...$ and arbitrary tau-function.

{\bf Remark 12.} It would be interesting to examine the functions $P_{\gamma}$ given by (\ref{g2}), (\ref{g2def}), (\ref{g2deg}) 
where $g=\infty,~E(x,y)=x-y,~q_i(z)=\frac{z^i}{i},~i=1,2,...$ and $\theta=\tau$. For example, (\ref{g2}) takes the form 
$$P_{\gamma}(z,u_1,...,u_n)=\int_{\gamma}\frac{\tau(\boldsymbol{\eta}+[z]-[t])}{(z-t)\tau(\boldsymbol{\eta})}\frac{(z-u_1)^{s_1}...(z-u_n)^{s_n}}
{(t-u_1)^{s_1}...(t-u_n)^{s_n}}e^{{\bf b}\cdot([z]-[t])}dt.$$ 
It particular, one could try to construct a Whitham type hierarchy (with infinitely many fields and times) associated with a universal moduli space containing all the 
Riemann surfaces of finite genus \cite{sch,orl}.

{\bf Remark 13.} It would be interesting to prove that Whitham type hierarchies constructed in this paper are integrable by hydrodynamic reductions for all genera and find corresponding 
Gibbons-Tsarev type systems \cite{Gibt}. 

These problems will be addressed in future publications.

\vskip.3cm
\noindent
{\bf Acknowledgments.}

I am grateful to E.V. Ferapontov, A.Yu. Orlov and V.V. Sokolov for useful discussions.

\end{document}